# Direct laser acceleration of electrons in free-space

Sergio Carbajo,[*,1,2,3] Emilio A. Nanni,[1] Liang Jie Wong,[1] R. J. Dwayne Miller,[3,4] Franz X. Kärtner[1,2,3]

[1]Department of Electrical Engineering and Computer Science, and Research Laboratory of Electronics, Massachusetts Institute of Technology, 77 Massachusetts Avenue, Cambridge, Massachusetts 02139, USA
[2]Center for Free-Electron Laser Science and The Hamburg Center for Ultrafast Imaging, Luruper Chaussee 149, 22607 Hamburg, Germany
[3]Deutsches Elektronen Synchrotron, and Department of Physics, University of Hamburg, Notkestraße 85, Hamburg 22761, Germany
[4]Max Planck Institute for the Structure and Dynamics of Matter, Luruper Chaussee 149, 22761 Hamburg, Germany and the Department of Chemistry and Physics, University of Toronto, 80 St. George Street, Toronto, Canada

**Compact laser-driven accelerators are versatile and powerful tools of unarguable relevance on societal grounds for the diverse purposes of science, health, security, and technology because they bring enormous practicality to state-of-the-art achievements of conventional radio-frequency accelerators[1]. Current benchmarking laser-based technologies[2-4] rely on a medium to assist the light-matter interaction, which impose material limitations or strongly inhomogeneous fields[5]. The advent of few cycle ultra-intense radially polarized lasers[6] has materialized an extensively studied[7-11] novel accelerator that adopts the simplest form of laser acceleration and is unique in requiring no medium to achieve strong longitudinal energy transfer directly from laser to particle. Here we present the first observation of direct longitudinal laser acceleration of non-relativistic electrons that undergo highly-directional multi-GeV/m accelerating gradients. This demonstration opens a new frontier for direct laser-driven particle acceleration capable of creating well collimated and relativistic attosecond electron bunches[10] and x-ray pulses[11].**

High-energy few-cycle laser sources with cylindrical vector beams remain relatively unexplored today, which has traditionally limited the generation of laser pulses with relativistic intensities to linearly polarized lasers. Radially polarized beams —a type of cylindrical vector beam— are especially of interest to relativistic laser-particle interactions because they exhibit cylindrical symmetry of the field-vector distribution and can therefore be focused much more tightly, down to about 0.6× the cross-sectional area of the diffraction-limited foci of linearly polarized beams[12,13]. As a consequence, the threshold power required to reach relativistic intensity levels becomes significantly less demanding. Moreover, the longitudinal field component of radially polarized focused beams is uniquely enhanced, with approximately twice the field-strength of beams with any other polarization, thereby making it the ideal driver for direct-field on-axis particle acceleration in vacuum[7,8,10,14].

Despite these unique properties and their promising theoretical performance in direct particle acceleration, no experimental evidence of direct laser acceleration of electrons in vacuum has been reported to date using cylindrical vector beams. The fundamental reason behind this shortcoming has up until recently lied upon the inability of several distinct methods[15-17] to scale up and deliver the minimum intensity requirements for few-cycle radially polarized beams to access the relativistic regime of laser-electron interaction, well above $10^{19}$ W/cm$^2$ in the case of optical pulses for instance. In fact, this technological void has given way to other laser acceleration schemes[18-24] today capable of yielding multi-GeV/m accelerating gradients and reaching well into the relativistic electron energy regime. However, these schemes require the presence of a medium for net energy transfer, which may translate into material breakdown or wave-breaking regime limitations for instance in the case of inverse Cherenkov acceleration[20] and plasma wakefield acceleration[25], respectively, or still face several challenges —large emittance, energy spread, and divergence[25,26] among others— which can severely hamper their use in practical applications. Direct acceleration employing radially polarized laser pulses is inherently a purely free-space scheme capable of creating well-collimated and relativistic attosecond electron bunches[10,11] and it is thereby in principle unrestrained from the limitations found in other methods.

Here we show an unprecedented experimental demonstration that serves as a proof of principle for direct in-vacuum longitudinal laser acceleration of electrons. This method takes advantage of the strong longitudinal electric field at beam center, where the transverse field components are not as strong, to accelerate electrons along the optical axis without additional mediation. The key enabling technology relies on a recently demonstrated efficient scheme to generate highly-focusable intense few-cycle radially polarized laser pulses[6]. We produce these pulses by combining a gas-filled hollow-waveguide compressor and a broadband linear-to-radial polarization converter with an ultrafast Ti:Sapphire chirp-pulse amplifier. The system delivers 3-cycle carrier-envelope-phase (CEP) stable radially polarized laser pulses at 3 kHz repetition rate centered at around 800 nm wavelength. The routinely operational peak- and average-power levels are 90 GW and 2.4 W, respectively. The system outputs very stable, high-quality, and nearly diffraction-limited TM$_{01}$-mode beams.

The quality of our few-cycle beams results in propagation factors close to unity and hence high focusability to access the relativistic intensity regime, defined as the region for which the ratio between the classical electron oscillation and the speed of light in vacuum —denoted as the normalized vector potential ($a_0$)— exceeds unity. Here, that ratio can reach as high as $a_0 \sim 5$. In this regime, non-relativistic electrons can pick up a significant amount of energy from the laser field and undergo gradients in the order of several tens of GeV/m, where the maximal net electron energy gain is proportional to the field intensity as well as the beam waist at focus[14,27]. Far from the center of the beam, electron in regions where the laser intensity is high enough will experience ponderomotive acceleration, while, near the center of the beam, the nature of the laser-electron interaction is twofold: (i) non-zero azimuthal angle electric field components may offer transverse confinement of the particles while (ii) they are accelerated (and decelerated) directly from the linear force exerted by the strong longitudinal component of the electric-field, uniquely available through cylindrically symmetric vector beams. In this study, we show first-time evidence of the laser-electron interaction mechanisms occurring near the center of the beam.

Our experimental setup (Fig. 1) captures the governing physics of longitudinal acceleration of non-relativistic electrons in vacuum. The 40 keV electron beam from the photocathode is 1 mm long and it is assumed to exhibit an even electron distribution over all phases of the laser field. The total bunch charge thereafter is 5 fC. The laser is focused with a high-NA parabolic mirror to a waist of 1.2 µm with a confocal parameter of 4 µm. The instrument is designed to limit the acceptance angle of electrons with less than 25 mrad azimuthal angle. We utilize a low-energy electron spectrometer consisting of magnetic dipole deflector (tangential plane) and a micro-channel plate with a nominal resolution of 100 eV/pxl.

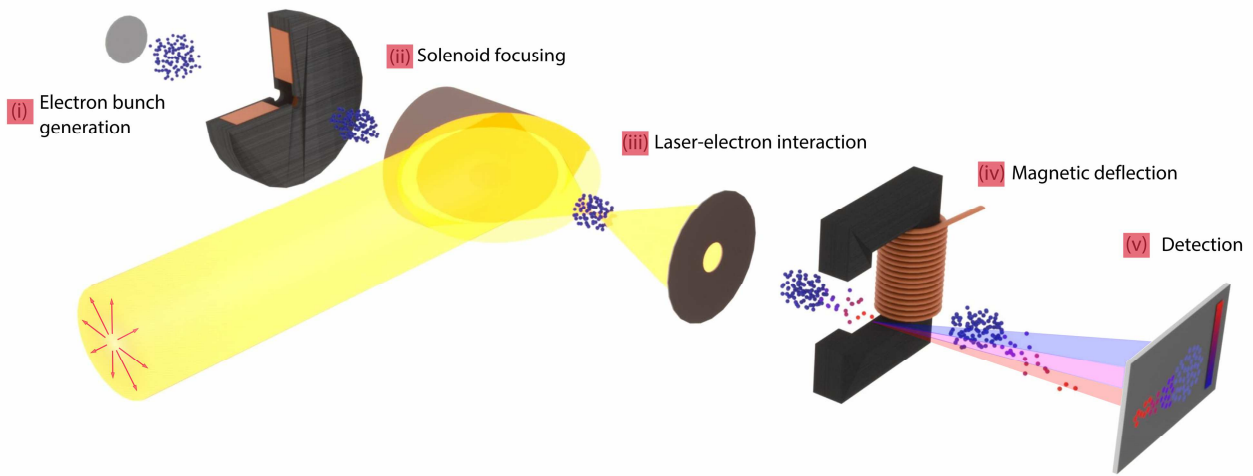

**Figure 1| Conceptual schematic.** A non-relativistic monoenergetic electron bunch is generated via a 40 kV DC gun (i) triggered by a small fraction of the optical beam. The electron bunch is then focused (ii) at the laser-electron IP using a solenoid. A subsequent off-axis parabola with a thru-hole allows for tight focusing of the optical beam to overlap with the electron bunch (iii). After the interaction, the electrons enter a dipole magnet deflector (iv) and are then mapped onto a CCD camera from the emission of a fluorescent screen placed after a micro-channel plate detector (v).

The exact spatiotemporal coincidence of the laser and electron beams are determined by photoionization-induced spatial distortions near the IP (more detail in Methods). We denote τ = 0 hereafter as to the relative time (τ) at which the accelerated charge peaks as a result of direct laser-electron interaction. Fig. 2.a shows that the time-window of this interaction is 5 ps wide, as expected from a millimeter-long electron bunch traveling at 0.39× the speed of light. The detected charge arises from accelerated electrons only in the deflection plane and does not exhibit any symmetry with respect to the laser field vector distribution. As a result, the direct interaction manifests itself as an increased energy spread of the electron beam with very small or negligible off-axis trajectory angle, where only the electrons close to the center of the focused beam are propelled forward. The divergence of the accelerated charge is less than 3 mrad full width at half maximum (FWHM), as shown in Fig.2.b. The relative strength of the acceleration and optimum overlap are depicted in real- and momentum-space at four distinct laser-electron overlapping times in Fig. 2.c. The electron bunch reveals a 5 mm transverse size at the detector plane.

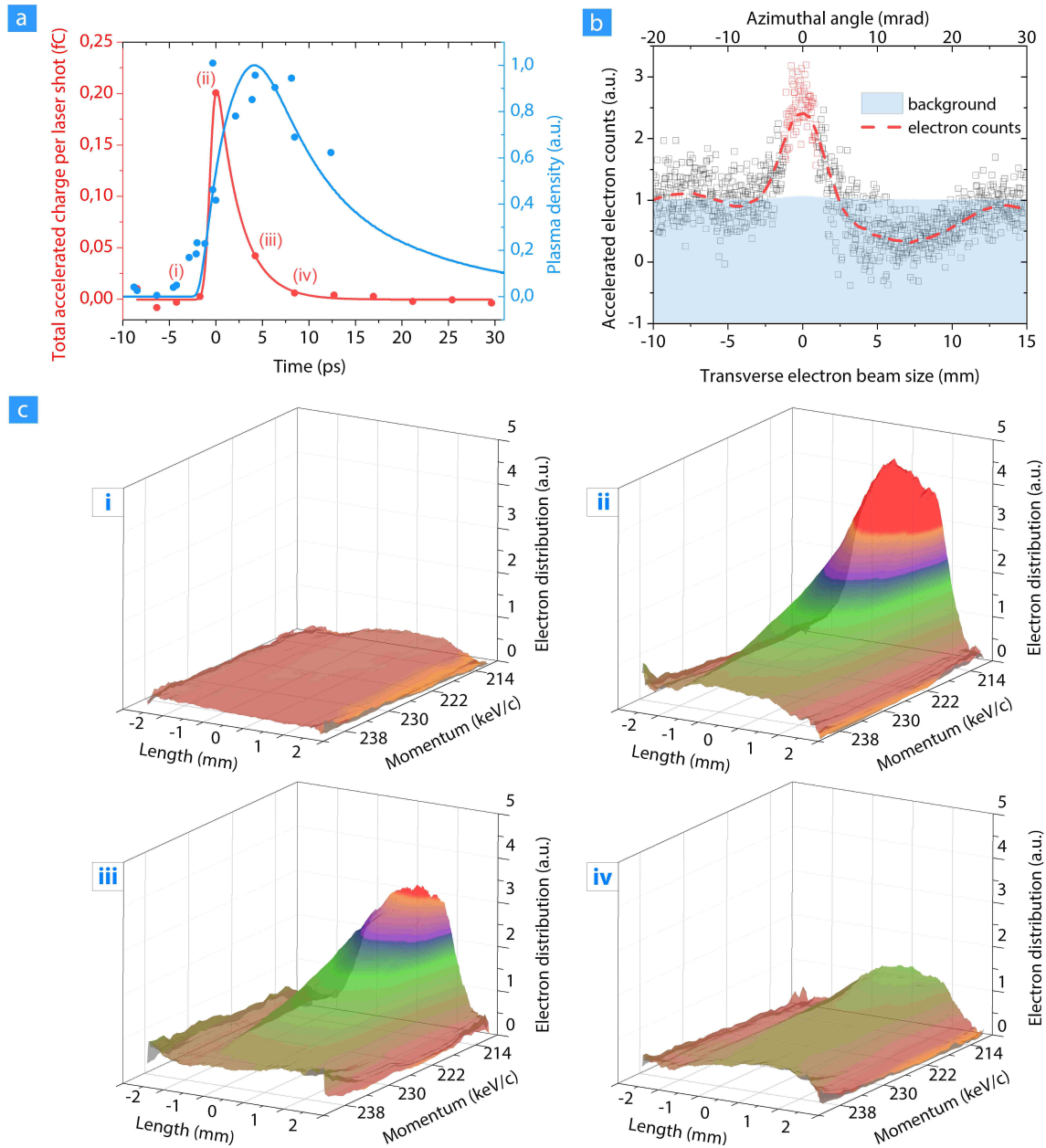

**Figure 2| Laser-electron interaction. a**, Electron-laser beam coincidence timing and normalized charge of detected accelerated electrons as a function of laser-electron timing delay (τ). **b**, Accelerated electron counts contained in the deflector plane at τ = 0. **c**, Snapshots of the normalized distribution in real- and momentum-space of accelerated electrons at four different temporal overlaps: **(i)** there are no accelerated electrons when the initial electron bunch arrives at the IP 8.5 ps before the laser; **(ii)** distribution at the peak of total accelerated charge (τ = 0) them beam is delayed by **(iii)** 4.25 ps and **(iv)** 8.5 ps with respect to the laser field.

In order to describe this interaction, we perform simulations based on a particle-tracking model beyond the paraxial and slowly varying envelope approximations. We model pulsed radially-polarized laser beams with exact, singularity-free solutions to Maxwell's equations[28,29]. We are able to fully recreate the spatial and momentum distribution of 4·10$^5$ particles anytime during and after collision. We restrict our attention to only an electron ellipsoid contained in the high intensity volume of the optical pulse. When the laser pulse reaches the significantly slower 40 keV electron bunch, the transverse portion in the center experiences acceleration from the largest longitudinal components of the laser field, which reaches its maximum on axis. Fig. 3 depicts the modeled and measured final energy distribution of such scenario. A significant portion of electrons further from the center may also be launched forward in phase ranges where the longitudinal fields are accelerating and transverse fields are focusing. This combined effect offers proper transverse confinement while accelerating particles along the beam axis.

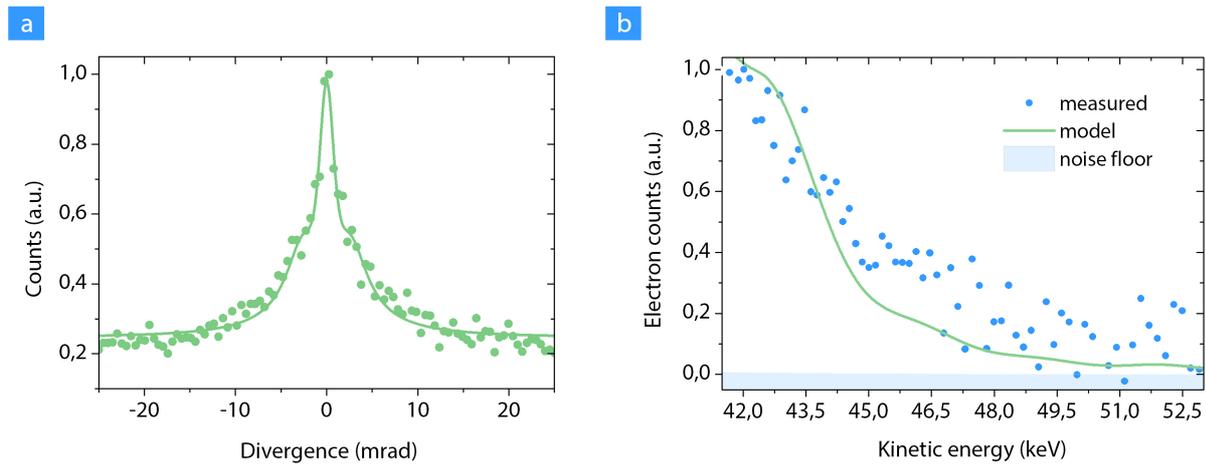

**Figure 3 | Electron bunch after interaction. a**, Modeled histogram of electrons with final electron kinetic energy ranging from 42 keV to 53 keV as a function of divergence with half-angle deflection of less than 25 mrad. From model, FWHM divergence of the electron bunch is expected to be 3.28 mrad. **b,** Final measured and modeled kinetic energy of accelerated electron in the region of interest up to 53 keV.

The predicted electron distribution matches the spatial (Fig.2.b) and energy (Fig.3.c) distribution measured at τ = 0, which exhibits very small angular spread and accounts for 4% of the total charge contained in the main electron beam, that is, 0.2 fC of accelerated charge per laser pulse. The net KE energy gain is strongest within a few keV and is measurable and statistically reproducible to up to 12 keV, thus comprising a 30% energy modulation of the initial electron KE, which is two orders of magnitude higher than that achieved by current state-of-the-art dielectric accelerators[21]. We demonstrate an unprecedented compact laser accelerator exhibiting highly-directional accelerating gradients up to 3 GeV/m at a minimum brilliance of 250 electrons/(s·mrad$^2$·mm$^2$·0.1%BW). Such gradients are comparable in order of magnitude to those achieved by current plasma-based accelerators[25], albeit at orders of magnitude less demanding energy levels from the driving laser pulses. In fact, employing petawatt-level few-cycle radially polarized lasers in free-space longitudinal acceleration would lead direct acceleration from rest to MeV-level[7].

This technology leverages the unrestricted scaling of usable laser power to inspire ultrafast high-repetition rate particle injectors and table-top medical radiation therapy treatment at a single-cell level, since it can readily access relativistic electron velocities when combined with radio-frequency injector technology. Alternatively, our demonstration is directly transferrable to any generalized form of direct longitudinal acceleration of charged-particles to higher energies and larger brilliance by upscaling the laser intensity and wavelength, respectively, and thus sets the stage for the development of compact atomic time- and space-scale resolution electron microscopy and tomography instruments with focusability to the Heisenberg limit, and portable high-brilliance tunable and coherent long-wavelength and attosecond x-ray sources.

**Acknowledgements**
This work was supported by DARPA under contract N66001-11-1-4192, by the Air Force Office of Scientific Research under grant AFOSR - A9550-12-1-0499, the Center for Free-Electron Laser Science at DESY and the excellence cluster "The Hamburg Centre for Ultrafast Imaging- Structure, Dynamics and Control of Matter at the Atomic Scale" of the Deutsche Forschungsgemeinschaft. The authors kindly acknowledge Gustavo Moriena and William S. Graves for their support and advice with experimental implementation.

**Author Contributions**
S. C., E. N., L. J. W., and F. X. K. are the conceptual authors of the work presented. S. C. and E. N. carried out the experimental results and L. J. W. performed the simulations. S. C. developed the laser source. E. N. and R. J. D. M. developed the electron beam instrumentation. All authors discussed the results and contributed to the final manuscript.

**References**

1. Malka, V. *et al.* Principles and applications of compact laser-plasma accelerators. *Nat Phys* **4**, 447-453 (2008).
2. Mangles, S. P. D. *et al.* Monoenergetic beams of relativistic electrons from intense laser-plasma interactions. *Nature* **431**, 535-538 (2004).
3. Toncian, T. *et al.* Ultrafast Laser-Driven Microlens to Focus and Energy-Select Mega-Electron Volt Protons. *Science* **312**, 410-413, doi:10.1126/science.1124412 (2006).
4. Tokita, S., Inoue, S., Masuno, S., Hashida, M. & Sakabe, S. Single-shot ultrafast electron diffraction with a laser-accelerated sub-MeV electron pulse. *Applied Physics Letters* **95**, -, doi:doi:http://dx.doi.org/10.1063/1.3226674 (2009).
5. Dwyer, J. R. *et al.* Femtosecond electron diffraction: 'making the molecular movie'. *Philosophical Transactions of the Royal Society A: Mathematical, Physical and Engineering Sciences* **364**, 741-778, doi:10.1098/rsta.2005.1735 (2006).
6. Carbajo, S. *et al.* Efficient generation of ultra-intense few-cycle radially polarized laser pulses. *Opt. Lett.* **39**, 2487-2490, doi:10.1364/OL.39.002487 (2014).
7. Wong, L. J. & Kärtner, F. X. Direct acceleration of an electron in infinite vacuum by a pulsed radially-polarized laser beam. *Optics Express* **18**, 25035-25051, doi:10.1364/OE.18.025035 (2010).
8. Pierre-Louis, F., Michel, P. & Charles, V. Direct-field electron acceleration with ultrafast radially polarized laser beams: scaling laws and optimization. *Journal of Physics B: Atomic, Molecular and Optical Physics* **43**, 025401 (2010).
9. Salamin, Y. I. Electron acceleration from rest in vacuum by an axicon Gaussian laser beam. *Physical Review A* **73**, 043402 (2006).
10. Varin, C. & Piché, M. Relativistic attosecond electron pulses from a free-space laser-acceleration scheme. *Physical Review E* **74**, 045602 (2006).
11. Sell, A. & Kärtner, F. X. Attosecond electron bunches accelerated and compressed by radially polarized laser pulses and soft-x-ray pulses from optical undulators. *Journal of Physics B: Atomic, Molecular and Optical Physics* **47**, 015601 (2014).
12. Dorn, R., Quabis, S. & Leuchs, G. Sharper Focus for a Radially Polarized Light Beam. *Physical Review Letters* **91**, 233901 (2003).
13. Kozawa, Y. & Sato, S. Sharper focal spot formed by higher-order radially polarized laser beams. *Journal of the Optical Society of America A* **24**, 1793-1798, doi:10.1364/JOSAA.24.001793 (2007).
14. Karmakar, A. & Pukhov, A. Collimated attosecond GeV electron bunches from ionization of high-Z material by radially polarized ultra-relativistic laser pulses. *Laser and Particle Beams* **25**, 371-377, doi:doi:10.1017/S0263034607000249 (2007).
15. Tidwell, S. C., Ford, D. H. & Kimura, W. D. Generating radially polarized beams interferometrically. *Appl. Opt.* **29**, 2234-2239, doi:10.1364/AO.29.002234 (1990).
16. Ahmed, M. A. *et al.* Radially polarized 3kW beam from a CO2 laser with an intracavity resonant grating mirror. *Opt. Lett.* **32**, 1824-1826, doi:10.1364/OL.32.001824 (2007).
17. Oron, R. *et al.* The formation of laser beams with pure azimuthal or radial polarization. *Applied Physics Letters* **77**, 3322-3324, doi:doi:http://dx.doi.org/10.1063/1.1327271 (2000).
18. Tajima, T. & Dawson, J. M. Laser Electron Accelerator. *Physical Review Letters* **43**, 267-270 (1979).
19. Buck, A. *et al.* Real-time observation of laser-driven electron acceleration. *Nat Phys* **7**, 543-548, doi:http://www.nature.com/nphys/journal/v7/n7/abs/nphys1942.html#supplementary-information (2011).
20. Kimura, W. D. *et al.* Laser Acceleration of Relativistic Electrons Using the Inverse Cherenkov Effect. *Physical Review Letters* **74**, 546-549 (1995).
21. Peralta, E. A. *et al.* Demonstration of electron acceleration in a laser-driven dielectric microstructure. *Nature* **503**, 91-94, doi:10.1038/nature12664 http://www.nature.com/nature/journal/v503/n7474/abs/nature12664.html#supplementary-information (2013).
22. van Steenbergen, A., Gallardo, J., Sandweiss, J. & Fang, J. M. Observation of Energy Gain at the BNL Inverse Free-Electron-Laser Accelerator. *Physical Review Letters* **77**, 2690-2693 (1996).
23. Mizuno, K., Pae, J., Nozokido, T. & Furuya, K. Experimental evidence of the inverse Smith-Purcell effect. *Nature* **328**, 45-47 (1987).
24. Plettner, T. *et al.* Visible-Laser Acceleration of Relativistic Electrons in a Semi-Infinite Vacuum. *Physical Review Letters* **95**, 134801 (2005).
25. Esarey, E., Schroeder, C. B. & Leemans, W. P. Physics of laser-driven plasma-based electron accelerators. *Reviews of Modern Physics* **81**, 1229-1285 (2009).
26. Geddes, C. G. R. *et al.* High-quality electron beams from a laser wakefield accelerator using plasma-channel guiding. *Nature* **431**, 538-541 (2004).
27. Carbajo, S., Wong, L. J., Nanni, E., Schimpf, D. N. & Kärtner, F. X. in *Research in Optical Sciences.* HTu2C.6 (Optical Society of America).

28  April, A. *Ultrashort, Strongly Focused Laser Pulses in Free Space*.  (INTECH Open Access Publisher, 2010).
29  Wong, L. J., Kärtner, F. X. & Johnson, S. G. Improved beam waist formula for ultrashort, tightly focused linearly, radially, and azimuthally polarized laser pulses in free space. *Opt. Lett.* **39**, 1258-1261, doi:10.1364/OL.39.001258 (2014).